\documentclass[pre,aps,twocolumn,showpacs,superscriptaddress]{revtex4} 
\usepackage{graphicx} 
\begin{document} 
\title{Ensemble Theory for Force Networks in Hyperstatic Granular 
Matter} 
 
\author{Jacco H. Snoeijer}  
\altaffiliation[Present address: ]{Physique et M\'ecanique des 
Milieux H\'et\'erog\`enes, ESPCI, 10 rue Vauquelin, 75231 Paris  
Cedex 05, France} 
\affiliation{Instituut--Lorentz, 
Universiteit Leiden, Postbus 9506, 2300 RA Leiden, The Netherlands} 
 
\author{Thijs J. H. Vlugt} \affiliation{Department of Condensed Matter 
and Interfaces, Debye Institute, Utrecht University, P.O.Box 80.000, 
3508 TA Utrecht, The Netherlands.} 
 
\author{Wouter G. Ellenbroek} \affiliation{Instituut--Lorentz, 
Universiteit Leiden, Postbus 9506, 2300 RA Leiden, The Netherlands} 
 
\author{Martin van Hecke} \affiliation{Kamerlingh Onnes Lab, 
Universiteit Leiden, Postbus 9504, 2300 RA Leiden, The Netherlands} 
 
\author{J. M. J. van Leeuwen} \affiliation{Instituut--Lorentz, 
Universiteit Leiden, Postbus 9506, 2300 RA Leiden, The Netherlands}

\date{\today} 
\begin{abstract} 
An ensemble approach for force networks in static granular 
packings is developed. The framework is based on the separation of 
packing and force scales, together with an a-priori flat measure 
in the force phase space under the constraints that the contact 
forces are repulsive and balance on every particle. In this paper 
we will give a general formulation of this force network ensemble, 
and derive the general expression for the force distribution 
$P(f)$. For small regular packings these probability densities are 
obtained in closed form, while for larger packings we present a 
systematic numerical analysis.  
Since technically the problem can be written as a non-invertible 
matrix problem (where the matrix is determined by the contact 
geometry), we study what happens if we perturb the packing matrix 
or replace it by a random matrix. The resulting $P(f)$'s differ 
significantly from those of normal packings, which touches upon 
the deep question of how network statistics is related to the 
underlying network structure. 
Overall, the ensemble formulation opens up a new perspective on force 
networks that is analytically accessible, and which may find applications 
beyond granular matter. 
\end{abstract} 
 
 
 
\maketitle 
 
\section{Introduction} 
 
One of the most fascinating aspects of granular media is the 
organization of the interparticle contact forces into highly 
heterogeneous force networks \cite{gm}. Direct evidence for these 
force networks mainly comes from numerical simulations \cite{radjai96,Pf} 
and experiments on packings of photo-elastic particles 
\cite{bob2d,shearclement}. While the contact physics can be quite 
convoluted \cite{contact}, numerical studies have shown that 
qualitatively similar force networks occur in systems with much 
simplified contact laws \cite{radjai96,Pf}. It has nevertheless remained a 
great challenge to understand the emergence of these networks and 
their properties. 
 
Even though the spatial structure and anisotropies of the force 
network may be important 
\cite{shearclement,vanel,2Dbehringer,radjai,PRL,herrmannanisotropy}, 
 a more basic quantity, the probability density of contact forces 
$P(f)$, has emerged as a key characterization of static granular 
matter 
\cite{network,lovol,edwards3D,radjai96,Pf,qmodel}. Recently this quantity 
has also been studied for a wider range of thermal and athermal 
systems \cite{edwards3D,liuletter}. Most of the attention 
so far has been focussed on the broad exponential-like tail of 
this distribution. Equally crucial is the generic change in 
qualitative behavior for small forces: $P(f)$ exhibits a peak at 
some finite value of $f$ for ``jammed'' systems which gives way to 
monotonic behavior above a glass transition 
\cite{liuletter,grestjam}. This hints at a possible connection 
between jamming, glassy behavior and force network statistics, and 
underscores the paramount importance of developing a theoretical 
framework for the statistics and spatial organization of the 
forces \cite{jamnote}. 
 
In this paper we study theoretical aspects of an ensemble approach 
that we recently introduced to describe these force networks 
\cite{PRL}. This force network ensemble is based on the separation 
of packing and force scales that occurs in systems of {\em hard 
particles}: in most experiments, typical grain deformations range 
from $10^{-2}$ to $10^{-6}$.  The crucial observation is that 
these packings are usually {\em hyperstatic}, i.e., the amount of 
force components is substantially larger than the number of 
force balance constraints \cite{grestcoord}.  This makes the 
problem ``underdetermined'' in the sense that there is no unique 
solution of the force network for a given packing configuration. 
For example, Fig.~1a shows two different force networks for a 
regular packing of 2D balls in a ``snooker-triangle''.  The 
ensemble is defined by assigning an {\em equal a priori 
probability to all force networks in which the net force on each 
particle is zero, for a given, fixed particle configuration}. 
Since we want to describe non-cohesive particles, we then consider 
only those networks that have purely repulsive forces. As can be 
seen from Fig.~1, 
these simple rules indeed yield configurations that resemble 
realistic force networks, as well as a force distribution $P(f)$ 
as typically observed in experiments and simulations. 
An important objective of this paper is to deepen our understanding 
of the force distribution, for this simplified but well-defined problem. 
 
\begin{figure}[tbp] 
\includegraphics[width=8.0cm]{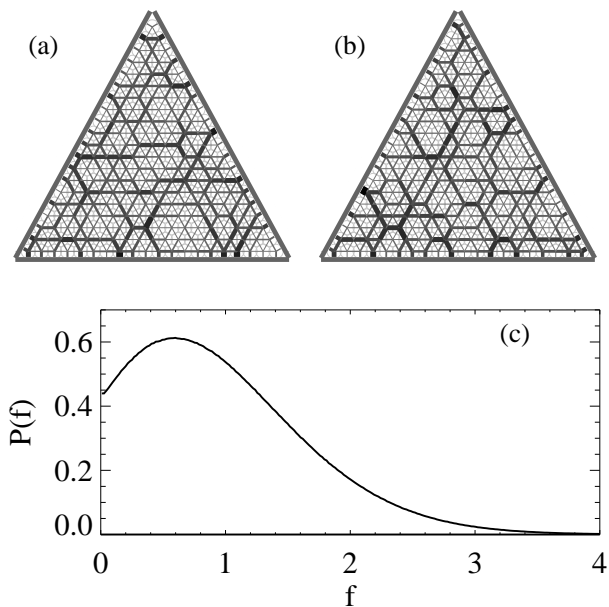} \centering 
\caption{(a--b) Two different mechanically stable force configurations 
for a ``snooker-triangle'' packing of 210 balls; the thickness of the 
lines is proportional to the contact force.  The ``force network 
ensemble'' samples all possible force configurations for a given 
contact network with an equal probability. (c) After sampling many 
force configurations, this yields the following distribution 
of interparticle forces $P(f)$.} \label{fig1} 
\end{figure} 
 
Our ensemble approach is in the same spirit of the Edwards 
ensemble, in which an equal probability for all ``blocked'' 
or ``jammed'' 
configurations is postulated \cite{edwards,kurchan}. This Edwards 
ensemble does not only average over forces, but also over all 
possible packing configurations, which makes the problem difficult 
to track theoretically. We therefore propose to exploit the 
separation of length scales that occurs for hard particles, by 
fixing the packing geometry (macroscopic scale) and allowing for 
force fluctuations (microscopic scale). Besides practical 
advantages, the conceptual gain of separating the contact geometry 
from the forces is that we can start to disentangle the separate 
roles of contact and stress anisotropies 
\cite{PRL,vanel,2Dbehringer}. Interestingly, the idea to restrict 
the Edwards ensemble to fixed packing geometry has also been 
proposed recently by Bouchaud in the context of extremely weak 
tapping \cite{jp}, and was also employed in recent simulations 
\cite{Unger,herrmannensemble}. 
Note that this force ensemble incorporates the local force balance 
equations on {\em all} particles and therefore it is fundamentally 
different from recent entropy-based models for force statistics 
\cite{ngan,max_entropy}. In these studies one postulates an entropy 
functional in terms of the single force distribution $P(f)$, 
without including the intricately coupled force balance equations 
and resulting force correlations.
 
From a more general point of view, the ensemble provides a 
challenging statistical physical problem of rather broad interest, 
that of sampling the solution space of a set of underdetermined 
equations and constraints. 
 For example, the problem is mathematically very similar 
to the so-called flux balance analysis that is used to unravel 
{\em metabolic networks} in biological systems 
\cite{nature,otherbio}. Here the reaction fluxes are 
underdetermined and play a role analogous to the forces discussed 
here. In contrast to the forces, however, these fluxes typically 
display power-law distributions \cite{nature}. This touches upon 
the deep question of what kind of statistics emerges when 
balancing scalars on a network of a given structure 
\cite{loadnetwork,qmodel}, and shows that the nature of the  
set of balance equations has a strong influence on the resulting  
statistics. 
 
The aim of this paper is to explore the ``phase space of force 
networks'' and to unravel how this gives rise to the robust 
characteristics of the force distribution $P(f)$.  We will 
initially focus on {\em regular packings} which are highly 
coordinated and therefore far from the isostatic limit. The 
advantage of these packings is, however, that the underlying  
physics is more transparent and that small regular 
packings can be resolved analytically. In addition, 
their force distributions are quite comparable to those found in 
numerical explorations of the ensemble for amorphous packings 
presented elsewhere \cite{PRL,longnumerics}. We will also study 
the ensemble on generalized networks, for which the force  
distributions rapidly loose their similarity to those of 
real packings. 
 
After defining the ensemble in more detail, the paper consists of 
four parts. In Sec.~\ref{sec.snooker}, we study the force ensemble 
for spherical, frictionless particles in regular triangular 
``snooker'' packings. We discuss how these force distributions are 
related to geometric aspects of the high-dimensional phase space. 
In Sec.~\ref{sec.general} we provide a formal mathematical 
description of the ensemble and derive the explicit form of 
$P(f)$, Eq.~(\ref{pfresult}).  This expression contains 
coefficients that depend on the packing geometry, and which we 
have been able to compute for several small systems. These exact 
$P(f)$ already exihibit the features that are relevant for larger, 
more realistic packings, and will be presented in 
Sec.~\ref{exactresults}.  Due to the linearity of the equations of 
force balance, the problem can be further generalized by 
considering perturbations of the packing matrix and random 
matrices, which are presented in Sec.~\ref{sec.rm}. This probes 
which ingredients are essential for obtaining realistic $P(f)$'s. 
The paper closes with a discussion of the strengths and weaknesses  
of our approach and indicates some open issues and  
other problems that can be addressed with the ensemble. 
 
\subsection{Definition of the force network ensemble} 
 
We will now introduce the main aspects of the ensemble approach. 
Even though our approach is perfectly suited to include frictional 
forces \cite{PRL,Unger,herrmannensemble}, for simplicity we will restrict 
ourselves to packings of $N$ frictionless spheres of radii $R_i$ with 
centers ${\bf r}_i$. We denote the interparticle force on particle $i$ 
due to its contact with particle $j$ by ${\bf f}_{ij}$. There are 
$zN/2$ contact forces in such packings ($z$ being the average contact 
number), and for purely repulsive central forces we can write ${\bf 
f}_{ij} = f_{ij} {\bf r}_{ij}/|{\bf r}_{ij}|$, where all $f_{ij} 
~(=f_{ji})$ are positive scalars. 
For a fixed contact topology in $d$ dimensions, 
we are thus left with $dN$ unknown positions ${\bf r}_i$ 
and $z N/2$ unknown forces $f_{ij}$. 
Note that the number of unknown forces is not 
precisely, but close to, $zN/2$ if boundary forces are present. 
 
These degrees of freedom satisfy the conditions of mechanical 
equilibrium, 
\begin{equation}\label{equilibrium} 
dN ~\mbox{eqs.:} \hspace*{3mm} \sum_j f_{ij}\,\frac{{\bf 
r}_{ij}}{|{\bf r}_{ij}|} ={\bf 0}, \hspace*{5mm}\mbox{where } {\bf 
r}_{ij} = {\bf r}_i-{\bf r}_j,\label{equilibria} 
\end{equation} 
and once a force law $F$ is given, the forces are explicit 
functions of the particle locations: 
\begin{equation} 
zN/2 ~\mbox{eqs.:} \hspace*{5mm} f_{ij} = F({\bf 
r}_{ij};R_i,R_j)~. \label{feqs} 
\end{equation} 
 
The contact number $z$ is a crucial quantity.  As has been argued 
before \cite{PRL,isostatic,isostatic2}, even though packings of 
infinitely hard frictionless particles have $z=2d$ and are thus 
{\em isostatic}, for particles of finite hardness, packings are 
typically {\em hyperstatic} with $z>2d$. In this paper we focus on 
hyperstatic packings, but before doing so, we wish to point out an 
important subtlety. In recent numerical work, it was shown that 
$z$ approaches the isostatic limit for vanishing pressures (hence 
vanishing deformations) of the particles, and that the (un)jamming 
transition here is similar to a phase transition, with power-law 
scaling of the relevant quantities and the occurrence of a large, 
possibly diverging lengthscale \cite{epitomy}. Therefore, the 
precise value of $z$ may be important, since it reflects the 
distance to the jamming phase transition; this may bear on the 
interpretation of our results.  It is worth pointing out that for 
frictional packings, even in the limit of infinitely hard 
particles, $z$ stays away from the isostatic limit 
\cite{grestcoord,kasahara}. Hyperstatic packings 
are therefore important, and our work, even though it focusses on 
frictionless packings, may also be seen in this light. 
 
Returning to the force network ensemble, in the regime where particles 
are hard but not infinitely hard, variations of the force of order 
$\langle f \rangle$ result in minute variations of ${\bf 
r}_{ij}$. Hence, Eqs.~(\ref{equilibria}) and (\ref{feqs}) can 
effectively be considered separated, and the essential physics is then 
given by the force balance constraints Eqs.~(\ref{equilibria}) with 
fixed ${\bf r}_{i}$. In this interpretation there are more degrees of 
freedom $(zN/2)$ than constraints $(2N)$, leading to an ensemble of 
force networks for a fixed contact geometry. 
 
This ensemble for a fixed contact geometry is then constructed as 
follows.  {\em{(i)}} Assume an a-priori flat measure in the force 
phase space $\{f\}$. {\em{(ii)}} Impose the $2N$ {\em linear} 
constraints given by the mechanical equilibrium 
Eqs.~(\ref{equilibria}). {\em{(iii)}} Consider repulsive forces only, 
i.e., $\forall f_{ij} \geq 0$.  {\em{(iv)}} Set an overall force scale 
by applying a fixed pressure or fixed boundary forces, similar to 
energy or particle number constraints in the usual thermodynamic 
ensembles. 
 
We are thus considering the phase space defined by the force balance 
Eqs.~(\ref{equilibria}), the condition that all $f$'s are positive, 
and a ``pressure'' constraint $\sum_k f_k = F_{\rm tot}$. 
For notational convience, we indicate the forces by a single index $k$ 
throughout the remainder of paper. 
Since all equations are linear, the problem can be formulated as 
\begin{equation}\label{fullmatrixproblem} 
{\cal A}\vec{f}= \vec{b} \quad\quad {\rm and} \quad \forall \quad 
f_k\geq 0~, 
\end{equation} 
where the fixed matrix ${\cal A}$ is determined by the packing geometry, 
$\vec{f}=(f_1, f_2,\cdots, f_{zN/2})$, 
and $\vec{b}=(0,0,0,\cdots, 0, F_{\rm tot})$.

\section{Regular packings: balls in a snooker triangle}\label{sec.snooker} 
 
\begin{figure}[tbp] 
\includegraphics[width=5.0cm]{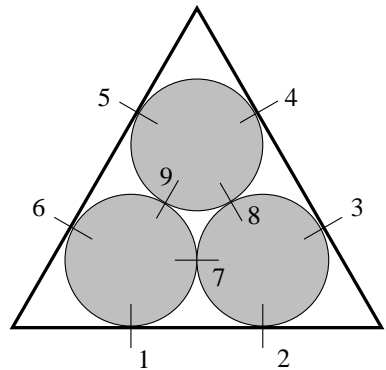} 
\centering 
\caption{Three monodisperse frictionless spheres in a snooker 
triangle.  This system has 9 unknown forces: 6 boundary forces ($f_1$ 
to $f_6$) and 3 interparticle forces ($f_7$, $f_8$, $f_9$).} 
\label{fig.3ballforces} 
\end{figure}

In the introduction we have seen that our ensemble approach for a 
snooker packing of 210 particles reproduces a force distribution 
that is very similar to those obtained in experiments and 
simulations.  To understand how this shape of $P(f)$ comes about, 
we now work out the force network ensemble for small systems of 
crystalline (monodisperse) packings.  We first study the packing 
of 3 balls shown in Fig.~\ref{fig.3ballforces}, for which we 
explicitly construct the phase space of force networks. As this 
system is very small, the force distribution deviates considerably 
from distributions observed in large systems. It nevertheless 
provides a very instructive example.  We then present a numerical 
analysis of how $P(f)$ evolves as a function of system size for 
snooker packings.  Remarkably, a packing of 6 balls is already 
sufficiently large to obtain the characteristic peak in $P(f)$. We 
therefore address general physical aspects by elaborating on this 
system.

\subsection{Three balls}\label{subsec.3balls} 
In the system of 3 balls depicted in Fig.~\ref{fig.3ballforces}, 
we encounter 9 unknown forces: 6 boundary forces and 
3 interparticle forces. These forces have to balance on 
each particle in both the $x$ and $y$ directions, which 
constitute $2\times3=6$ linear constraints. 
In addition, we impose an overall pressure by keeping 
the total force on a boundary at a fixed value: 
for example we fix $f_1+f_2=2$. Interestingly, one can show 
that such a boundary or pressure constraint is equivalent 
to keeping the sum over {\em all} forces at a fixed value: 
in appendix \ref{app1} we demonstrate that 
keeping $\sum_j f_j$ at a constant value 
is equivalent to a constant pressure, 
also for irregular packings.

\begin{figure}[tbp] 
\includegraphics[width=8.0cm]{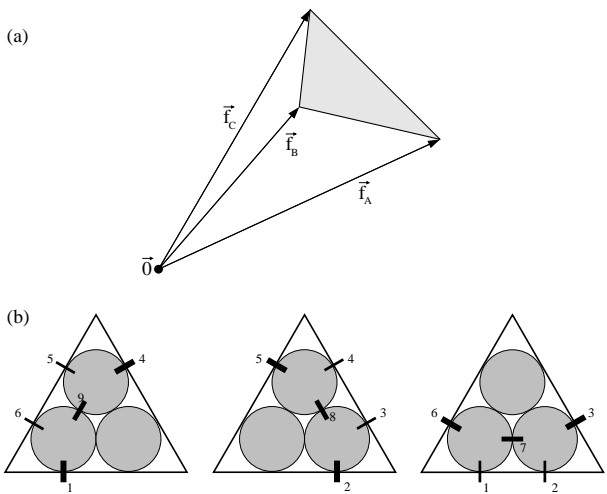} 
\centering 
\caption{{\em (a)} The 2D phase space of the 3-ball problem can be defined 
using three simple independent solutions of the problem. 
{\em (b)} The first solution $\vec{f}_A$ has $f_1=f_4=2$, $f_5=f_6=1$ and 
$f_9=\sqrt{3}$ 
and $f_{2,3,7,8}=0$; the other solutions $\vec{f}_B$ and $\vec{f}_C$ 
follow from the threefold symmetry of the packing.} 
\label{fig.3ballmode} 
\end{figure} 

Together with the pressure constraint, there are thus 7 linear 
equations to determine the 9 unknown forces $\vec{f}=(f_1;, \cdots, f_9)$, 
and hence, there is a two-dimensional space of solutions. 
This space does not contain the origin of the force space, 
for which all $f_j=0$, due to the inhomogeneous pressure constraint. 
As a consequence one requires three vectors to 
characterize the two-dimensional space: 
two basis vectors and a vector defining the location of 
the plane with respect to the origin. 
Using linear algebra one can construct these three vectors from 
three linearly independent force network solutions 
$\vec{f}_A$, $\vec{f}_B$, $\vec{f}_C$, which allows to 
express the general solution as

\begin{equation} 
\vec{f}=c_A \vec{f}_A + c_B \vec{f}_B + (1-c_A -c_B) \vec{f}_C~.
\label{3ballsf} 
\end{equation} 
An intuitive picture of this equation is provided in 
Fig.~\ref{fig.3ballmode}a: 
the 2-dimensional plane can be defined from three solutions 
(very much like a line can be defined by two points). 
However, the constraints that all $f_j\geq 0$, provide serious 
limitations on the allowed values of $c_A$ and $c_B$. 
As will be shown below, only a small convex subset of the 
the two-dimensional solution space represents force networks 
consisting of strickly repulsive forces.

\begin{figure}[tbp] 
\includegraphics[width=8.0cm]{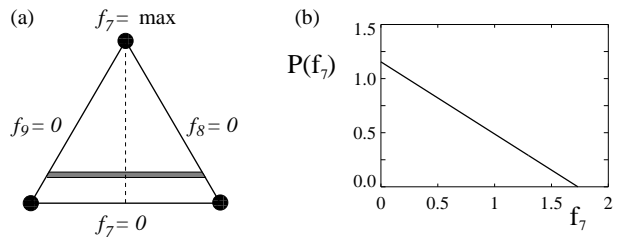} 
\centering \caption{Two dimensional cut through the phase-space 
spanned by the nine forces of the 3-ball problem. {\em (a)} The 
borders of the triangle are the lines where one of the 
interparticle forces changes sign; the shaded area represents the 
probability to find a configuration between $f_7$ and $f_7 + 
\delta f_7$. {\em (b)} The corresponding force distribution 
$P(f_7)$.} \label{fig.3ballspace} 
\end{figure} 
 
Using the solutions of Fig.~\ref{fig.3ballmode} to construct the phase 
space, we obtain the triangle depicted in 
Fig.~\ref{fig.3ballspace}a. In this picture, the three solutions are 
the corners of the triangle.  For example, the right corner represents 
the first solution in Fig.~\ref{fig.3ballmode}, $\vec{f}_A$ 
for which $(f_7,f_8,f_9)=(0,0,\sqrt{3})$, 
whereas the left corner corresponds to 
$\vec{f}_B$ that has $(f_7,f_8,f_9)=(0,\sqrt{3},0)$.  A superposition of 
these two vectors is still a solution of our linear problem, and since 
in both cases $f_7=0$, the base of the triangle is a line where 
$f_7=0$. The upper corner represents $\vec{f}_C$, for which $f_7$ 
attains its maximum value of $\sqrt{3}$.  Therefore, the dashed line 
is a projection of the $f_7$ axis onto this 2D space of 
solutions. This implies that the space below the triangle corresponds 
to a region where $f_7 < 0$, which is forbidden for repulsive 
particles. Applying the same argument for $f_8$ and $f_9$, one 
realizes that only the area inside the triangle is allowed.  As we 
mentioned in the introduction, the ensemble assumes an equal a priori 
force probability, which makes each point in the triangle equally 
likely (due to the linearity of the force balance restrictions). 
Therefore, the probability to have a 
solution between $f_7$ and $f_7+\delta f_7$ is simply represented by 
the shaded area in Fig.~\ref{fig.3ballspace}a.  This ``volume'' 
decreases linearly as $f_7$ approaches its maximum value, so that the 
distribution of $f_7$ simply becomes $P(f_7)=\frac{2}{3} (\sqrt{3}-f_7) 
\Theta(\sqrt{3}-f_7)$ -- see Fig.~\ref{fig.3ballspace}b. 
 
The combination $x\Theta(x)$, where $\Theta(x)$ is the Heaviside step  
function, will occur in most $P(f)$ thoughout this paper.  
Therefore we introduce 
\def\calt{\mathcal{T}} 
\begin{equation} 
\label{defcalT} 
\calt(x)\equiv x\Theta(x)~. 
\end{equation} 
 
\begin{figure}[tbp] 
\includegraphics[width=8.0cm]{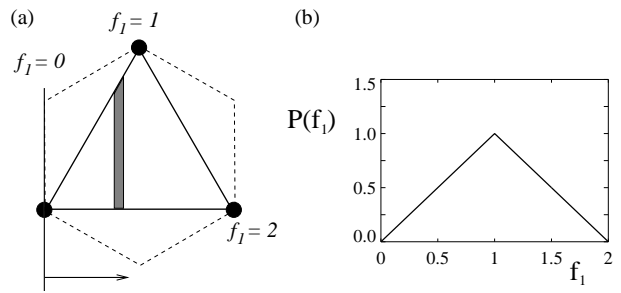} 
\centering \caption{Two dimensional cut through the phase-space 
spanned by the nine forces of the 3-ball problem, showing how 
boundary forces are distributed. {\em (a)} The borders of the 
hexagon are the lines where one of the boundary forces changes 
sign; the shaded areas represents the probability for a certain 
$f_1$. {\em (b)} The corresponding force distribution $P(f_1)$.} 
\label{fig.3ballboundary} 
\end{figure} 
 
The distribution of the boundary forces ($f_1$ to $f_6$) can be 
found in a similar manner. Checking the three independent 
solutions, one finds that $f_1=0$ at the left corner, $f_1=1$ at 
the upper corner, and $f_1=2$ at the right corner of the ``phase 
space triangle''.  From the geometric construction in 
Fig.~\ref{fig.3ballboundary}a, it is easy to find the $f_1=0$ 
line, and the projection of the $f_1$ axis is indicated by the 
arrow. Due to symmetry there are of course six such borders 
($f_1=0$ to $f_6=0$), and all boundary forces are positive inside 
the hexagon. So, the solutions for which {\em all} forces are 
positive lie within the triangle.  Considering the shaded area in 
Fig.~\ref{fig.3ballboundary}a, we obtain the distribution of boundary 
forces $P(f_1)= \calt(2-f_1) - 2\calt(1-f_1)$, which is 
shown in Fig.~\ref{fig.3ballboundary}b. 
We thus find that there is a qualitative difference between the 
boundary forces $(f_1,\cdots,f_6)$ and the interparticle forces 
$(f_7,f_8,f_9)$. Interestingly this is also the case for larger 
systems and is consistent with earlier work on statistics  
of wall forces \cite{rapid,loooong}. 
 
Although this 3-ball system provides a very nice illustration of how 
to obtain $P(f)$ from all possible force configurations, it is not 
complex enough to reproduce non-monotonic $P(f)$.  In fact, the 
problem discussed above is equivalent to partitioning a conserved 
energy into 3 positive parts.  In our case, the conserved quantity is 
the total force and the 3 parts are the coefficients $c_A$, $c_B$ and 
$(1-c_A-c_B)$.  In the thermodynamic limit, the problem of 
partitioning, e.g., an energy $E_{\rm tot}$ simply yields the 
Boltzmann distribution of energies $E_i$ of subsystems; also for 
finite systems these distributions are always monotonically decreasing 
-- see appendix~\ref{app1}.  In Sec.~\ref{subsec.6balls}, we show that 
the problem of 6 balls already has enough complexity that it leads to 
non-monotonic behavior of $P(f)$. 

\subsection{Numerical analysis of larger systems}\label{sect.numsnooker} 
 
\begin{figure}[tbp] 
\includegraphics[width=8.0cm]{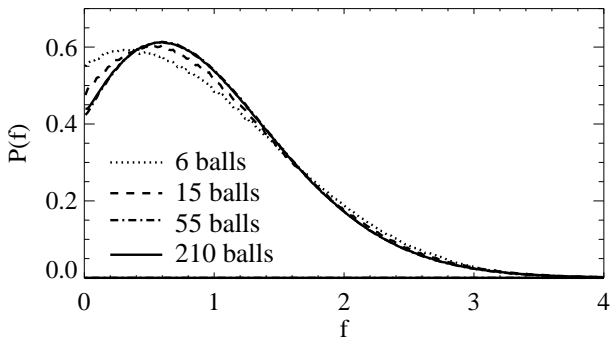} \centering \caption{$P(f)$ for 
bulk forces in ``snooker'' packings of increasing sizes.} 
\label{fig.fig2} 
\end{figure} 
 
To compute $P(f)$ for larger packings, we have applied a simulated 
annealing procedure \cite{numrec}. 
As was shown in our previous work \cite{PRL} this scheme can also be 
used for irregular packings. 
Starting from an ensemble of random initial force
configurations taken from an arbitrary distribution with $\langle f
\rangle = 1$ and $f_j \geq 0$, we select a random bond $j$ and add a
random force $\Delta f$, so that $f_j(n) = f_j(o) + \Delta f$, 
in which the symbols $n$ and $o$ denote the new and old
force respectively. The random change from $o$ to $n$ is accepted 
with a probability given by the conventional Metropolis rule 
$p(o \rightarrow n) = \min \left(1, \theta \left(f_j(n)\right) \exp
\left[-\left({\cal H}(n) - {\cal H}(o) \right)/T \right]  \right)$, 
in which $\cal{H}$ is a penalty function whose degenerate ground
states are solutions of Eq.~(\ref{fullmatrixproblem}):
\begin{equation}
{\cal H}(\vec{f})= \left( {\cal A}\vec{f} -  \vec{b} \right)^2 .
\label{eq:pentaltyh}
\end{equation}
For large packings 
($N>500$) it is computationally much more efficient to always
satisfy $\langle f \rangle = 1$ 
by selecting two bonds ($j\neq k$) at random and using 
$f_j(n) = f_j(o) + \Delta f$ and $f_k(n) = f_k(o) - \Delta f$ 
as the update scheme, so that the pressure constraint can be 
left out of the penalty function. By slowly taking the limit of
$T \rightarrow 0$ we sample all mechanically stable force
configurations for which ${\cal H} \rightarrow 0$. We have 
carefully checked that results do not depend on the initial 
configurations and details of the annealing scheme. 
In section \ref{exactresults} we will show that this scheme perfectly 
reproduces analytic results for small regular packings. 
 
The two force networks shown in Fig.~\ref{fig1} are typical solutions 
$\vec{f}$ obtained by this procedure. The resulting distributions 
of interparticle forces are presented in Fig.~\ref{fig.fig2}, 
for packings of increasing number of balls; 
boundary forces will be discussed seperately and are not 
included in these $P(f)$. 
Note that all $P(f)$'s display a peak for 
small $f$, which is typical for jammed systems \cite{liuletter}. 
The fact that the probability for vanishing interparicle forces 
remains finite is in agreement with most numerical and experimental 
observations; only a few studies report power-law behavior for small 
forces \cite{radjai96,lovol}. 
For large packings, this peak rapidly converges to its asymptotic limit. 
The tail 
of $P(f)$ broadens with system size, and will be discussed in more detail 
in Sec.~\ref{sec.discussion}. 
 
\begin{figure}[tbp] 
\includegraphics[width=8.0cm]{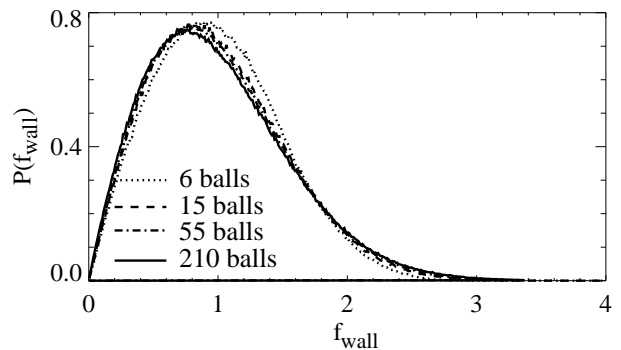} \centering 
\caption{Boundary forces for ``snooker'' packings of increasing 
sizes.} 
\label{fig_bound} 
\end{figure} 
 
\def\fwall{f_\mathrm{wall}} 
In Fig.~\ref{fig_bound} we show the probability distributions $P(\fwall)$ 
for the forces $\fwall$ between sidewalls and balls for the regular 
packings of increasing size. As has been discussed at length in 
\cite{rapid,loooong}, these distributions differ from the probability 
distributions of bulk forces. In this particular case this is easy to 
see: the boundary force $\fwall$ has to balance the force of the two balls 
in the next layer with which it makes contact (excluding the corner 
balls). Even though each of these forces has a finite probability to 
be vanishingly small, the probability that {\em both} these forces 
are small has not, hence $P(\fwall)\rightarrow 0$ for $\fwall \rightarrow 0$.

\subsection{$P(f)$ and phase space geometry}\label{subsec.6balls}

\begin{figure}[tbp] 
\includegraphics[width=8.0cm]{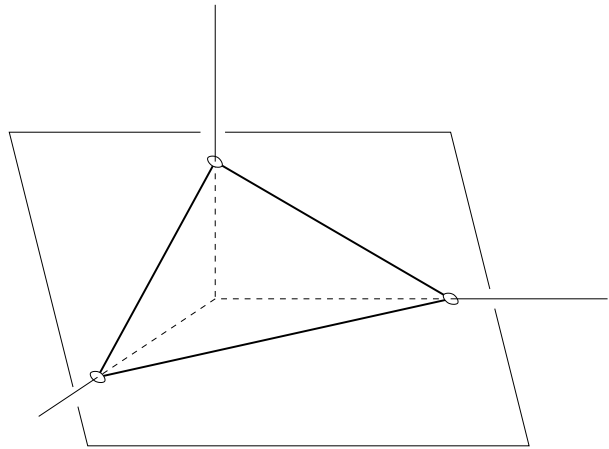} \centering 
\caption{Schematic representation of the phase space of allowed force 
configurations. Each $f_{ij}$ defines a direction in the $zN/2$ 
dimensional force space. By imposing the linear conditions of 
mechanical equilibrium, this space is restricted to a ``hyperplane'' of 
lower dimensionality. The physically allowed region is a 
(hyper)polygon is bounded by the requirement that all $f_{ij}\geq0$.} 
\label{fig2} 
\end{figure}

Here we will discuss some geometrical aspects of the set of allowed 
force configurations. Consider the $zN/2$ dimensional force phase 
space spanned by the ${f_j}$. Since all $2N$ force balance equations 
(\ref{equilibrium}) are {\em linear} in the forces, the allowed 
solutions lie on a hyperplane of dimension $(zN/2-2N)$. (Note that the 
overall pressure constraint introduces an additional constraint, 
lowering the dimension by 1.)  Furthermore, since we consider 
repulsive forces only, this plane is restricted to the ``positive'' 
hyperquadrant where all $f_j \geq 0$ (see Fig.~\ref{fig2}). Therefore, the 
allowed force-configurations form a (hyper)-polygon whose {\em 
facets} are given by the conditions that some force $f_j$ becomes 
0. Under our assumption of a ``flat measure'', all points on this 
polygon correspond to valid force networks with equal probability. 
 
A number of basic properties of this solution space can now readily be 
deduced. Trivially, the solution space is convex: due to  the linearity 
of the equations, the points on a straight line connecting two admissible 
solutions are admissible solutions themselves, as was also pointed out in 
\cite{bioconvex,Unger,herrmannensemble}.  
Although this is not immediately obvious in low dimensions, for higher 
dimensional bodies the overwhelming part of the ``measure'' is 
concentrated near the boundary (think of a high dimensional sphere, 
where almost all volume is in the ``shell'' close to surface). Near 
the boundaries, one or more forces tend to zero, and this is consistent 
with the fact that in typical force networks a finite fraction of the 
forces are close to zero (since $P(f\downarrow0) \neq 0$).  More 
homogeneous force networks, for which {\em all} forces are around some 
average value, correspond to points in the phase space that are 
sufficiently far away from the boundary. While such configurations are 
perfectly allowed within our framework, and are easy to construct by 
considering a suitable linear combination of ``ordinary'' 
force-networks, they only occur with vanishingly small probability in 
the limit of large $N$, and are thus extremely unlikely to be seen in 
``unguided numerics'' or experiments. 
 
Even though we have not worked this out in detail, we expect that some 
more general properties of the force networks could be related to 
geometrical properties of (random) hyperpolygons. As one simple 
example consider the following. For two forces, say $f_i$ and $f_j$, 
to become zero simultaneously, the facets $i$ and $j$ have to touch; 
in general this may not be possible geometrically, so that an 
intruiging issue concerning correlations between distant forces arises. 
 
Another issue that may have a relatively simple interpretation in the 
polygon language is the peaked appearance of $P(f)$. We suggest the 
following intuitive picture, based on a consideration why the slope 
$dP(f)/df$ can be expected to be positive for small forces. For very 
small systems, like the case of 3 balls discussed in section 
\ref{subsec.3balls}, this is not true. This immediately follows from the shape 
of the allowed phase space polygon. As shown in 
Fig.~\ref{fig.3ballspace}, this is a triangle where the angles between 
the bounding edges were {\em acute}. When we move away from a $f=0$ 
boundary, the phase space volume decreases so that $dP/df <0$. If we go to 
larger systems, however, the number of facets bounding the space 
$(=zN/2)$ becomes much larger than the dimension $D$ $(=zN/2-dN-1)$ 
of the polygon. Hence, we 
expect that the ``angles'' between bounding facets will typically 
become {\em obtuse}, which will make the phase space {\em increase} 
when increasing $f$. This indicates that $dP/df$ is typically positive 
for small forces, so that $P(f)$ displays a peak \cite{footfacets}. 
 
\begin{figure}[tbp] 
\includegraphics[width=8.0cm]{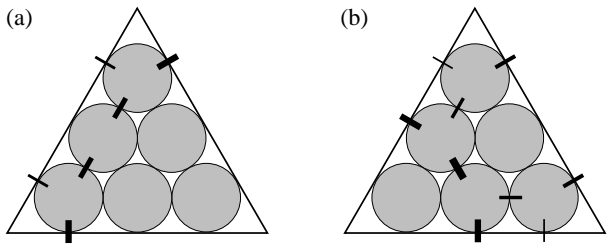} 
\centering 
\caption{Two different types of solutions of force 
equilibrium for 6 balls in a snooker-triangle.} 
\label{fig.6ballforces} 
\end{figure}

{\em Six balls} -- 
Let us provide another perspective on the phase space 
geometry by discussing the problem of 6 balls, which is the 
smallest snooker packing displaying a non-monotonic $P(f)$. 
For the 6 balls there are $18$ forces, which are constrained 
by $2\times 6 +1=13$ equations, so the space of solutions is 
a 5D hyperplane. If we try to construct the phase space 
like we did for the 3 balls, we now require 6 
independent solutions $\vec{f}$ that obey force balance 
on each particle. Again, there exist simple solutions of 
linearly propagating force lines - see Fig.~\ref{fig.6ballforces}a. 
However, there are only 3 such solutions, so we also require 
nontrivial solutions where forces ``scatter'' at a certain particle. 
For example, we can take 3 solutions of the type shown in 
Fig.~\ref{fig.6ballforces}b.

The presence of these nontrivial solutions changes the phase space 
in a a fundamental manner. 
A given force can now take a certain value in many different ways, 
by different linear combinations of elementary ``modes''. 
In other words, a force can no longer be associated to a single 
mode of the force network, like it was the case for the three 
interparticle forces in Fig.~\ref{fig.3ballmode}.  
As a consequence, the problem has become much more intricate than 
simply partitioning the total force into positive amplitudes 
(which, for large systems, would lead to a simple exponential 
distribution, see appendix~\ref{app1}). Instead one finds nontrivial force distributions, 
for which we derive analytical expressions in the following 
section. Indeed, for all investigated packings, we observe 
non-monotonic $P(f)$ whenever scatter-solutions occur.

\section{General formulation for arbitrary packing geometry}\label{sec.general} 
 
In this section we show how statistical averages can be computed 
analytically within the force network ensemble, for arbitrary 
packings. We present a systematic way to evaluate the complicated 
high-dimensional integrals as a sum over contributions of the 
following structure:

\begin{eqnarray}\label{pfresult} 
P(f)&=&\sum_\lambda  c_\lambda\, f^{q_\lambda} 
(1-b_{\lambda} f)^{D-1-q_\lambda}\Theta (1-b_\lambda f )\nonumber\\ 
&=&\sum_\lambda  c_\lambda\, f^{q_\lambda} 
\left[\calt(1-b_\lambda f)\right]^{D-1-q_\lambda}~, 
\end{eqnarray} 
where $D$ is the dimension of the phase space, and the coefficients 
$b_\lambda$, $c_\lambda$ and $q_\lambda$ depend in a nontrivial way on 
the particle packing; for most $\lambda$, we find that 
$q_\lambda=0$. The function $\calt$ was defined in 
Eq.~(\ref{defcalT}); note that the contributions 
$\left[\calt(1-b f)\right]^{D-1} \sim e^{-(D-1)bf}$ in the 
thermodynamic limit. For the reader 
who is interested in the results but not in the details of the 
derivation, we summarize exact $P(f)$ for small regular packings in 
Sec.~\ref{exactresults}. 
 
\subsection{Mathematical definition of the ensemble} 
The phase space of force networks is defined by the linear 
constraints of force balance, an inhomogeneous linear constraint to fix the 
pressure, and the requirement that all forces are non-negative. 
If we now indicate each contact force by an index $j$, 
we can express mechanical equilibrium as 
 
\begin{equation}\label{matrixmechequilibrium} 
\sum_{j=1}^{zN/2} a_{ij} f_j = 0~, 
\end{equation} 
where the nonzero $a_{ij}$ are projection factors between 
$-1$ and $+1$. There are $dN$ such equations, which we label 
as $i=2,3,\cdots, dN+1$. To keep the overall pressure 
at a fixed value we impose $\sum_j f_j = F$, 
which for notational convience we write as 
 
\begin{equation}\label{pressure} 
\sum_{j=1}^{zN/2} a_{1j} f_j = F~, \quad\quad \mbox{\rm with all} 
\quad a_{1j}=1~. 
\end{equation} 
We thus encounter a matrix problem ${\cal A}\vec{f}=\vec{b}$, 
where the $a_{ij}$ are the components of ${\cal A}$. 
 
Imposing the various constraints and assuming an 
{\em equal a priori probability} in the force space defined 
by $\vec{f}=(f_1,f_2,\cdots,f_{zN/2})$, 
we obtain the joint probability density 
 
\begin{eqnarray} 
P(\vec{f}) &=& \frac{1}{\Omega}\, 
\delta\left( \sum_j a_{1j} f_j - F \right) 
\prod_{i\geq2} \delta\left( \sum_j a_{ij} f_j \right)~, 
\end{eqnarray} 
which is normalized by the phase space volume 
 
\begin{equation}\label{volume} 
\Omega = \int d\vec{f} \,\delta\left( \sum_j a_{1j} f_j - F \right) 
\prod_{i\geq2} \delta\left( \sum_j a_{ij} f_j \right)~. 
\end{equation} 
Since we consider repulsive forces, $\int d\vec{f}$ represents 
an integral over all forces in the hyperquadrant where all $f_j\geq0$. 
With this measure, we can now compute the single force 
distribution $P(f_j)$ as 
 
\begin{equation}\label{defpf} 
P(f_j) = \left[\prod_{k\neq j}\int_0^\infty df_k\right] \, P(\vec{f}) \,~, 
\end{equation} 
which in principle can be different for each $f_j$; for example see 
the boundary forces within the snooker-triangles (Sec.~\ref{sec.snooker}). 
In practice, it turns out that $P(f_j)$ for different 
{\em interparticle} forces shows only little variation. 
 
The fact that we only integrate over the hyperquadrant where 
all $f_j\geq0$ makes it difficult to evaluate the integrals 
explicitly: each integration of the $\delta$ function gives rise 
to a Heaviside $\Theta$ function to keep track of the boundaries 
of the phase space. To avoid this problem we represent the $\delta$ 
functions as Fourier integrals 
 
\begin{equation}\label{fourierdelta} 
\delta\left(\sum_j a_{ij}f_j\right) = 
\int_{-\infty}^\infty \frac{ds_i}{2\pi} \, e^{-is_i \sum_j a_{ij}f_j}~, 
\end{equation} 
which has the advantage that the $f_j$ only occur in an 
exponential way and they are easily integrated out. 
If we now write $\vec{s}=\frac{1}{2\pi}(s_1,\cdots,s_m)$, 
where $m=dN+1$, the partition function $\Omega$ becomes 
 
\begin{eqnarray}\label{partitionfunction} 
\Omega&=&\int_{-\infty}^\infty d\vec{s} \,e^{is_1F} \, 
\prod_j \int_0^\infty df_j \, 
e^{-(\epsilon_j + i\sum_i s_i a_{ij})f_j} \nonumber \\ 
&=& \int_{-\infty}^\infty d\vec{s} \, e^{is_1F} \, 
\prod_j \frac{1}{i(-i\epsilon_j+\sum_i s_i a_{ij})}~, 
\end{eqnarray} 
where the factor $e^{is_1F}$ arises due to the inhomogeneous 
pressure constraint (\ref{pressure}). 
We furthermore added cut-off factors $e^{-\epsilon_j}$ so that the 
integrations over the $f_j$ are definite; 
at the final stage we take the limit $\epsilon_j \rightarrow 0$. 
The rows of the matrix ${\cal A}$ correspond to the constraint 
variables $s_i$ and the columns correspond to the denominators 
originating from the $f_j$ integrals. From now on we indicate 
the dimensions of the matrix by $m=dN+1$ (number of rows) 
and $n=zN/2$ (number of columns). 
 
All integration variables $s_i$ run from $-\infty$ to $\infty$, 
so we can evaluate them as contour integrations in the complex plane. 
The integrand is a product of denominators, and each $s_i$ occurs 
in as many denominators as there are forces acting on a certain 
particle. In the absence of gravity, each mechanically stable 
particle should at least have $3$ contacts. This makes the integration 
over the $s_i$ converging at infinity and allows to close the contour 
either through the upper half plane or through the lower half plane. 
An exception is the $s_1$ integration, which has to be closed 
through the upper plane since $F>0$. 
 
Let us first integrate out $s_m$. Each denominator that has 
$a_{mj}\neq 0$ gives rise to a pole at 
 
\begin{equation} 
s_{m}(j)=\frac{1}{a_{mj}} 
\left(i\epsilon_j - \sum_{i=1}^{m-1} s_i a_{ij}\right)~. 
\end{equation} 
The residue is obtained by substiting this pole in the 
remaining $n-1$ denominators of Eq.~(\ref{partitionfunction}). 
Note the importance of the $\epsilon_j$ to make the integration definite. 
It is easily seen that this 
substitution leads to a renormalized matrix ${\cal A}^*$ of $m-1$ 
rows (constraint variables) and $n-1$ columns (denominators), 
and to renormalized $\epsilon_{j'\neq j}^*$ as well. 
However, the key observation is that the remaining 
integrals are of the same type as Eq.~(\ref{partitionfunction}). 
We thus find a recursion relation 
 
\begin{equation}\label{recursionomega} 
\Omega_{mn}({\cal A})=\pm \sum_j \frac{1}{a_{mj}} \, 
\Omega_{m-1,n-1}({\cal A}^*_j)~, 
\end{equation} 
where the sum extends over all encircled poles. 
The symbol $\pm$ indicates that the contribution is positive or 
negative depending on whether the integral has been closed through 
the upper ($+$) or lower ($-$) half plane. 
The renormalization to ${\cal A}^*_j$ is 
different for each pole, so each term has to be followed 
independently. At each integration the number of contributions 
therefore grows rapidly, since each new pole gives rise to a new 
``branch'' of the recursion Eq.~(\ref{recursionomega}).  
The exponential increase of the number of branches with the size 
of the tree forms a severe limitation on the solutions for larger systems. 
At the final stage, we have to compute 
$\Omega_{1,n_{\rm final}}=\Omega_{1,D+1}$ by 
integrating over $s_1$: 
 
\begin{eqnarray}\label{omegafinal} 
\Omega_{1,D+1} &=& 
\int_{-\infty}^\infty \frac{ds_1}{2\pi} \, e^{is_1F} \, 
\prod_j \frac{1}{i(-i\epsilon_j+ s_1 a_{1j})} \nonumber \\ 
&=& \frac{F^{D}}{D!\prod_j a_{1j}}~, 
\end{eqnarray} 
where $D$ is the dimensionality of the phase space. 
The $a_{1j}$ and ${\epsilon_j}$ appearing in this equation 
are obtained from successive renormalization each time 
a pole is substituted. 
 
So, the calculation of $\Omega$ involves a tree-like structure 
where the branching rate is equal to the number 
of encircled poles. Using relation Eqs.~(\ref{recursionomega}) 
and~(\ref{omegafinal}) one can compute the contribution 
of each individual branch, using a recursive scheme. 
The fact that $\Omega$ scales as $F$ to the power $D$ is not surprising: 
$F$ is the only force scale for the $D$ dimensional phase space, 
and in fact, the behavior $F^D$ is obtained immediately from a 
trivial rescaling of Eq.~(\ref{volume}). However, in the following 
paragraphs we show how the analysis presented above can be extended 
to the nontrivial calculation of the force distribution $P(f)$. 
 
\subsection{Calculation of $P(f)$} 
Comparing Eqs.~(\ref{volume}) and~(\ref{defpf}), we notice that 
the expression for $P(f_j)$ is the same as that for $\Omega$ without 
the integration over $f_j$; without loss of generality we will 
consider $P(f_1)$. As a result, the expression for $P(f_1)$ contains 
one less dominator than Eq.~(\ref{partitionfunction}) 
and instead there will be an additional exponential factor, i.e. 
 
\begin{eqnarray}\label{pfstart} 
P(f_1)&=&\frac{1}{\Omega} 
\int d\vec{s} \, e^{i\left(s_1F -f_1\sum_i s_i a_{i1}\right)} 
\, e^{-\epsilon_1 f_1} \nonumber \\ 
&&\times \, \prod_{j\neq1} \frac{1}{i(-i\epsilon_j+\sum_i s_i a_{ij})}~. 
\end{eqnarray} 
Following the same integration strategy as for $\Omega$, we again 
obtain a recursion of the type 
\begin{equation}\label{recursionpf} 
P_{m,n}(f_1)=\pm \sum_k \frac{1}{a_{mk}} \, 
P_{m-1,n-1}(f_1)~, 
\end{equation} 
where for clarity in notation we left out the explicit 
dependence on the (renormalized) matrix ${\cal A}$. 
After successive substitution of the poles, 
the final integration over $s_1$ becomes 
 
\begin{eqnarray}\label{pffinal} 
P_{1,D+1}(f_1) &=& \frac{1}{\Omega} 
\int_{-\infty}^\infty \frac{ds_1}{2\pi} \, 
e^{is_1\left(F-a_{11}f_1\right)} 
\, e^{-\epsilon_1 f_1} \nonumber \\ 
&&\times \, \prod_{j\neq1} 
\frac{1}{i(-i\epsilon_j+ s_1 a_{1j})} \nonumber \\ 
&=& \frac{1}{\Omega}\, 
\frac{\left( F-a_{11}f_1 \right)^{D-1}}{(D-1)!\prod_{j\neq 1} a_{1j}} 
\, \Theta\left(  F-a_{11}f_1\right)~. \nonumber \\ 
&& 
\end{eqnarray} 
Each branch of the tree gives a contribution of this type 
and together they accumulate to the result of Eq.~(\ref{pfresult}) 
with $q_\lambda=0$. 
The coefficients $b_\lambda$ are thus simply the $a_{11}/F$ that remain 
after successive renormalization of the matrix ${\cal A}$. 
We will demonstrate that, fortunately, the final result contains only 
a few different $b_\lambda$, at least for small packing geometries. 
 
In the final integration of Eq.~(\ref{pffinal}), we implicitly 
assumed that all $a_{1j}$ appearing in the denominators are 
not equal to zero. They may become negative, provided that the 
associated small $\epsilon_j$ is also negative so that 
the pole is still in the upper half plane and the integration 
remains finite. 
Naively one would expect that it very unlikely that some $a_{1j}=0$, 
since it corresponds to an accidental coincidence of two poles. 
However, for regular structures like the snooker-packings 
it is a frequently occuring phenomenon.  
The double poles are responsible for the cases $q_\lambda\neq 0$.  
We have adapted the algorithm such that it can deal with an arbitrary  
multiplicity of the poles. 
In some cases, these multiple poles alter the general result for 
$P(f)$ with additional contributions of the type 
 
\begin{equation}\label{multiplepoles} 
P_\lambda(f) \propto f^{q_\lambda}\left(1-b_\lambda f \right)^{D-1-q_\lambda} 
\,\Theta\left(1-b_\lambda f \right)~. 
\end{equation} 
These contributions can be recognized as the $q_\lambda$th derivatives of 
the general result, corresponding to the coincidence of $q_\lambda+1$ poles. 
We expect, however, that multiple poles will never occur for 
disordered packings.

\section{Exact results for small crystalline packings}\label{exactresults} 
We now present a number of exact $P(f)$ for small crystalline packing 
geometries. In particular, we have worked out the problem of 6 balls 
in a snooker-triangle, triangular 2D packings with periodic boundary 
conditions, as well as a small 3D FCC packing with 
periodic boundary conditions.  Following the algorithm described in 
the previous section, we have been able to obtain the coefficients 
$b_\lambda$ and $c_\lambda$ appearing in Eq.~(\ref{pfresult}) for 
these systems.  For notational convenience, we express the results in 
the dimensionless force $x=f/F$.  All is in perfect agreement with our 
numerical simulations. 
 
The intricate combinatorics has been performed using a 
computer program.  
As mentioned the number of contributions grows exponentially with the size 
of the tree, since the branching rate is of order of 2 per elimination step. 
Even worse is the fact that the different signs of the contributions lead to large 
cancellations. The results given below for small systems are the result of 
many more terms in the tree. This makes the algorithm numerically unstable for 
larger systems. 
 
\subsection{2D triangular packings with periodic boundaries} 
 
{\em Four balls} -- 
The smallest interesting 2D triangular packing with periodic 
boundary conditions is the $2\times2$ packing of 4 balls. 
It has $3\times 4=12$ unknown forces and $2\times 4=8$ 
equations expressing mechanical equilibrium. 
Due to the periodic boundaries, however, two of these equations 
are actually dependent. Hence there are only $6$ independent equations 
and together with the overall pressure constraint this results 
into a $D=12-(6+1)=5$ dimensional phase space. 
 
\begin{figure}[tbp] 
\includegraphics[width=8.0cm]{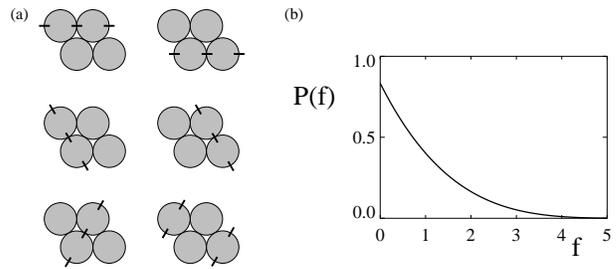} 
\centering 
\caption{{\em (a)} All solutions of the $2\times2$ periodic 
arrangement can be described as a superposition of 
linearly propagating force-lines. {\em (b)} The corresponding 
monotonic $P(f)$.} 
\label{fig.4ballsperiodic} 
\end{figure}

In terms  of the dimensionless variable $x=f/F$, we obtained the 
following result for this system: 
 
\begin{equation}\label{px2bij2} 
P(x)= 10\, \calt^4(1-2 x)~. 
\end{equation} 
Taking $F=12$ so that $\langle f \rangle=1$, 
we plotted this distribution in Fig.~\ref{fig.4ballsperiodic}b. 
It is a monotonically decreasing function that allows a maximum force of 
$x_{\rm max}=\frac{1}{2}$,i.e., $f_{\rm max}=\frac{1}{2} F$. 
This maximum force is achieved for a simple ``propagating'' 
solution shown in Fig.~\ref{fig.4ballsperiodic}a: 
the total force $F$ is shared between two nonzero forces only 
(note the similarity to the solutions shown in Fig.~\ref{fig.3ballmode} 
for the packing of three balls). Due to the symmetry of the 
problem there are six such trivial solutions, which are in 
fact sufficient to define the whole 5D phase space of 
force networks. The $2\times2$ problem is therefore equivalent 
to partitioning the total force into six nonnegative 
``amplitudes'', just as was the case for the three balls in the 
snooker-triangle. Indeed, Eq.~(\ref{px2bij2}) is of the 
same form as Eq.~(\ref{px}) in appendix~\ref{app1}.

\begin{figure}[tbp] 
\includegraphics[width=8.0cm]{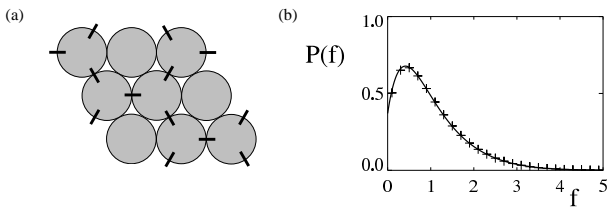} 
\centering 
\caption{{\em (a)} The system of $3\times 3$ balls 
allows for nontrivial ``scatter'' solutions. {\em (b)} The 
corresponding $P(f)$ is therefore non-monotonic. The solid 
curve is Eq.~\ref{px3bij3}; the crosses are obtained from 
numerics as described in section \ref{sect.numsnooker}.} 
\label{fig.9ballsperiodic} 
\end{figure} 
 
{\em Nine balls} -- 
For the $3\times3$ packing of 9 balls there are 
$3\times9=27$ unknown forces that are constrained 
by $2\times9-2=16$ independent equations of mechanical equilibrium. 
Fixing the overall pressure, one is left with a 
$D=27-(16+1)=10$ dimensional phase space. 
This space can {\em not} be reconstructed from the 
trivial propagating solutions, of which there are only 9. 
Again, the presence of the ``scatter'' solutions such as the one shown 
in Fig.~\ref{fig.9ballsperiodic}a results into a non-monotonic $P(x)$: 
 
\begin{eqnarray}\label{px3bij3} 
P(x) = 40\left[ \calt^9(1-3 x) - \frac{3}{4} \calt^9(1-9 x)\right]~. 
\end{eqnarray} 
Taking $F=27$ so that $\langle f \rangle=1$, we plotted $P(f)$ as a solid curve in 
Fig.~\ref{fig.9ballsperiodic}b; the crosses indicate the distribution obtained by 
the same numerical method that was used for the snooker-triangles in 
Sec.~\ref{sec.snooker}.  The perfect agreement illustrates the accuracy of our 
numerical method.

\subsection{3D FCC packing with periodic boundaries} 
 
To illustrate that our ensemble can be applied to three 
dimensional packings just as well,  
we have computed $P(f)$ in the conventional FCC unit cell, with 
periodic boundary conditions. This is a system of 4 balls, since the FCC unit 
cell contains $8$ particles at corners (each counting for $1/8$) and 
$6$ particles at the faces (counting for $1/2$).  
The coordination number of the FCC packing is $z=12$, 
so there are $zN/2=24$ forces in this system. 
We now have to respect force balance in three dimensions, 
i.e. $3\times 4=12$ equations, of which, due to 
periodic boundary conditions, only $9$ turn out to be independent. 
Together with the pressure constraint, there are thus $10$ 
equations to constrain $24$ forces, and hence the problem has 
a $14$ dimensional space of solutions. 
 
The resulting $P(x)$ turns out to be 
 
\begin{eqnarray} 
P(x)&=& \frac{364}{9}\left[\calt^{13}(1-2x)-\frac{9}{26}\calt^{13}(1-6x) \right. \nonumber \\ 
\label{Pf-FCC}&& \left.-\frac{4}{13}\calt^{13}(1-8x) -27x\calt^{12}(1-6x)\right]~. 
\end{eqnarray} 
 
\begin{figure}[tbp] 
\includegraphics[width=8.0cm]{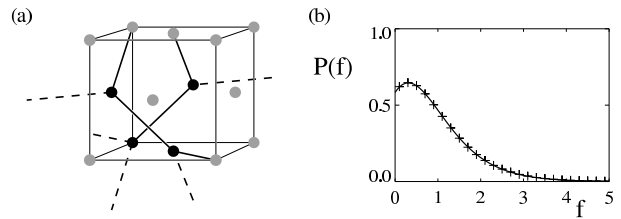} 
\centering 
\caption{{\em (a)} One of the ``scatter'' solutions for 
the FCC unit cell with periodic boundary conditions. The black 
spheres belong to this unit cell; the grey spheres belong to neighbouring cells. 
All forces have the same magnitude; those within this unit cell are drawn as thick solid lines; 
the others are drawn as thick dashed lines. 
{\em (b)} The corresponding non-monotonic $P(f)$, from Eq.~\ref{Pf-FCC} (solid curve) and from 
numerics as described in section \ref{sect.numsnooker} (crosses).} 
\label{fig.fccperiodic} 
\end{figure} 
Fig.~\ref{fig.fccperiodic} shows that this force distribution has the same typical 
features as those obtained for two-dimensional packings.  
It's a non-monotonic function, which can again be related to the 
existence of ``scatter'' solutions. There are 15 independent solutions 
to fix the 14D phase space of force networks, 12 of which are linearly 
propagating ``trivial'' solutions (two for each lattice direction). The 
other three are again ``scatter'' solutions. 
One of these is shown in Fig.~\ref{fig.fccperiodic}.

\subsection{6 balls in a snooker-triangle} 
 
We now provide the exact force distributions for the 6 balls in a 
snooker-triangle, which we discussed in Sec.~\ref{sec.snooker}. We 
already showed that one has to distinguish between the {\em 
interparticle forces} and the {\em particle-wall forces}, which obey 
qualitatively different statistics. Upon closer inspection, however, 
one notices that there are also two different types of interparticle 
force: the 6 closest to the boundary (type $I$) and the 3 closest to 
the center (type $II$). We find that 
 
\begin{eqnarray} 
P_I(x) = 
\frac{95 a^5}{768\left( 7+4\sqrt{3}\right)} 
\times \nonumber \\ 
\left[  \calt^4(1-ax) -\frac{16}{19}\calt^4(1-2ax) + 
\frac{3}{19}\calt^4(1-3ax) \right]~, 
\end{eqnarray} 
 
\begin{eqnarray} 
P_{II}(x) = 
\frac{15 a^5}{64\left(7+4\sqrt{3} \right)} 
\times \nonumber \\ 
\left[  \calt^4(1-ax) 
-\frac{5}{6}\calt^4\left(1-\frac{3}{2}ax\right) 
- ax\calt^3\left(1-\frac{3}{2}ax\right) \right]~. 
\end{eqnarray} 
where $a=2\left( 1+\sqrt{3}\right)$. 
 
The numerical results shown in Fig.~\ref{fig.fig2} were obtained 
without discriminating between type $I$ and type $II$.  This is 
allowed since even though $P_{I}(x)$ and $P_{II}(x)$ are not 
identical, their shapes are very similar. Comparing the data with 
$\frac{2}{3} P_{I}(x)+ \frac{1}{3} P_{II}(x)$ gives again an excellent 
agreement between the theoretical result and the numerical result 
shown in Fig.~\ref{fig.fig2}. 
The factors 2/3 and 1/3 appear because there are 6 forces of 
type $I$ and 3 forces of type $II$.
 
Finally, let us discuss the statistics for the boundary forces as 
shown in Fig.~\ref{fig_bound}. Also in this case there are two 
different types of boundary forces, namely 6 at the corners $(c)$ and 
3 in the middle $(m)$ of each boundary. We find that 
 
\begin{eqnarray} 
P_{c}(x) &=& \frac{5 b^5}{9\left(7+4\sqrt{3} \right)} 
\left[\calt^4(1-bx) -\calt^4(1-2bx) \right. \nonumber \\ 
&& \left. -\frac{10}{3}bx\calt^3(1-2bx)- 2(bx)^2\calt^2(1-2bx) \right]~, \nonumber \\ 
&& 
\end{eqnarray} 
 
\begin{eqnarray} 
P_{m}(x) = \frac{5 b^5}{54\left(7+4\sqrt{3} \right)} 
\times \nonumber \\ 
\left[-\calt^4(1-bx) + 8bx\calt^3(1-bx) + \calt^4(1-2bx) \right]~, 
\end{eqnarray} 
where $b=3 + \sqrt{3}$. The linear combination $\frac{2}{3} P_{c}(x)+ 
\frac{1}{3} P_{m}(x)$ fits the boundary force distributions as shown 
in Fig.~\ref{fig_bound} extremely well (not shown).

\section{Beyond Packings}\label{sec.rm} 
 
In the preceding sections we have extensively studied the force 
distributions emerging in the ensemble of force networks, for a 
variety of crystalline packings. The various $P(f)$ are 
non-monotonic and display only marginal differences. As we 
demonstrated in Ref.~\cite{PRL}, the same qualitative behavior is 
observed for irregular packings. Even though the packing matrices 
differ substantially in these cases, the resulting $P(f)$ is 
extremely robust. This raises the question of which are the 
essential ingredients to obtain a typical force distribution. In 
other words, what properties of the packing matrix ${\cal A}$  
determine the shape of $P(f)$? 
 
All packing matrices consist of a large number of zeros, except 
for a few elements per row that are projection factors between 
$-1$ and $1$. Such a matrix has some features of a random matrix, 
but it implicitly contains the entire spatial structure of the 
system. To see whether this spatial structure is crucial for the 
typical shape of $P(f)$, we now study true {\em random matrices}, 
which no longer represent a physical packing of particles. Of 
course, we still extend the matrix by the normalization constraint 
$\sum_j f_j=F$ and demand that all $f_j\geq 0$. 
 
We find that such random matrices yield $P(f)$ whose decay 
is described by a product of Gaussian and exponential tails. 
However, all these distributions are monotonically decreasing 
and thus lack the typical peak, 
even when considering ``sparse'' random matrices. 
We then try the opposite approach, where we start from a 
physical packing matrix and then slowly introduce randomness. 
In contrast to the striking robustness of $P(f)$ for real 
packings, the force distribution is very sensitive even 
to small perturbations away from the physical matrix.

\subsection{Random matrices} 
 
\subsubsection{Infinite Gaussian random matrices} 
 
We start out the random matrix approach by generating all 
elements $a_{ij}$ in Eq.~(\ref{matrixmechequilibrium}) 
from a Gaussian distribution 
 
\begin{equation} 
P_a(a_{ij}) = \sqrt{\frac{1}{\pi}} \, 
e^{-a_{ij}^2}~, 
\end{equation} 
for which the problem can be solved exactly. 
Together with the constraint $\sum_j f_j=F$, 
we obtain a matrix of $m$ rows and $n$ columns. 
By demanding that all $f_j\geq 0$, one can in principle 
follow the same analysis as for real packings; 
we then average over all possible 
random matrices and consider only solutions with all $f_j\geq0$. 
In appendix~\ref{app2} we derive that, 
in the limit that $n,m\rightarrow \infty$ with a fixed ratio 
$\rho = m/n$, the distribution becomes 
\begin{equation}\label{gaussmatrixresult} 
P(f) = c(\rho) \, e^{-(1-\rho) f} \, e^{-b(\rho)f^2}~, 
\end{equation} 
where $b(\rho)$ is an almost linear function that has  
$b(0)=0$ and $b(1)=1/\pi$. 
For square matrices, i.e. $\rho=1$, we thus find that $P(f)$ is a 
pure Gaussian centered around $f=0$. This is illustrated in 
Fig.~\ref{fig.randmat1}a; to calculate the $P(f)$ for these 
non-square matrices we have evaluated Eqs.~(\ref{pf}) and~(\ref{defpeff})  
by Monte Carlo simulation. Tuning $\rho$ to 
zero, the pressure constraint is dominant and we retrieve the pure 
exponential behavior that is also discussed in 
appendix~\ref{app1}. 
 
So, we find that the tail of $P(f)$ is a mixture of a Gaussian and an 
exponential, depending on the aspect ratio $\rho$ of the matrix. 
However, for any value of $\rho$ it is monotonically decreasing, 
and we never observe the peak that is extremely robust for real 
packing matrices. 
 
A relevant question of course is whether a Gaussian distribution 
of all matrix elements is representative for a matrix that is 
based on a real system of particles. Such a ``real'' packing matrix 
is not only sparse but also has $a_{ij}\in[-1,1]$ in such a way 
that Newton's third law is respected. Unfortunately, it becomes 
very hard to work out the integrations if $P_a(a_{ij})$ is not 
Gaussian \cite{footlorentz} or when correlations between matrix 
elements are imposed. For those systems, we have to rely on 
numerical simulations. 
 
\subsubsection{Numerical simulations}\label{randomnumerics} 
 
To numerically sample the ensemble discussed above, one first has 
to average over a representative number of allowed $\vec{f}$ for 
each matrix ${\cal A}$, and then repeat this for many different 
matrices. However, only very few of the generated matrices 
actually have solutions for which all $f_j \geq0$. We have 
therefore focused our study on {\em square} random matrices, for 
which the phase space consists of a single point and the numerical 
effort is thus reduced to inverting each matrix. Starting from a 
random matrix for which all $f_j \geq0$, we apply a Monte Carlo 
simulation procedure in which attempts are made to change a 
randomly selected element of ${\cal A}$ (except the elements 
corresponding to the pressure constraint). 
Such attempts are accepted with a probability given by 
the conventional Metropolis acceptance/rejection rules \cite{fre022}. In this 
way, we are able to explore the phase space of random matrices for 
which all $f_j \geq0$, for any distribution of the matrix elements $a_{ij}$. 
 
It is important to note that this numerical procedure is not 
precisely equivalent to the analysis of the Gaussian  
random matrices presented above.  
The reason for this is that a {\em flat} or {\em uniform} 
measure is not uniquely defined for continuous variables: a 
nonlinear transformation of variables gives rise to a Jacobian 
that affects this flat phase space density. Since the coupling 
between $a_{ij}$ and $f_j$ is indeed nonlinear, the flat measure 
is ambiguous. However, one can show that the measure of the 
numerical scheme differs by a factor $\det({\cal A})$ from 
$P\left(\vec{f},{\cal A} \right)$ of Eq.~(\ref{measure}), and we 
have verified that including this ``weight factor'' in the 
simulations only mildly alters the $P(f)$ for small matrices ($n 
\leq 5$) and practically disappears for larger matrices. 
 
\begin{figure}[tbp] 
\includegraphics[width=8.0cm]{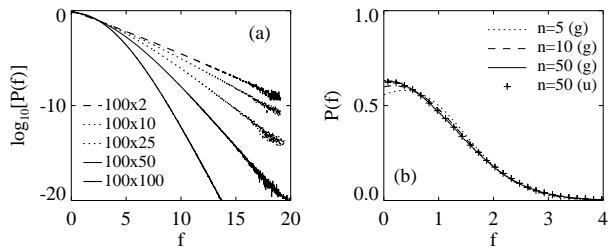} \centering 
\caption{(a) Numerical evaluation of $P(f)$ for matrices with dimensions  
ranging from $100 \times 2$ $(\rho \approx 0)$ to $100 \times 100$ 
($\rho = 1$) illustrating crossover from exponential to Gaussian 
behavior (compare to Eq.~\ref{gaussmatrixresult}). (b) Force 
distributions obtained with $n\times n$ Gaussian random matrices 
(with pressure constraint) for different values of $n$ (curves). 
For $n=50$ the force distribution obtained with matrix elements 
from a \emph{uniform} distribution is included for comparison 
(crosses).} \label{fig.randmat1} 
\end{figure} 
 
{\em Square random matrices} -- 
Let us start the discussion with $n\times n$ square random matrices 
like the ones used for the analytical calculation above. This means one 
of the rows of the matrix represents the pressure constraint and 
the others are taken from a Gaussian distribution. 
In the limit $n\rightarrow \infty$ these were shown to give rise to 
a (half) Gaussian force distribution, 
see Eq.~(\ref{gaussmatrixresult}) with $\rho=1$. 
The numerical results for $n=5,10,50$ are shown in 
Fig.~\ref{fig.randmat1}b. The distribution for $n=50$ is indeed a 
Gaussian, as expected for $n\to\infty$. The case $n=5$ displays a 
very small peak at finite $f$, but this effect disappears quickly 
when $n$ increases. Furthermore, Fig.~\ref{fig.randmat1}b shows that the 
distribution obtained with Gaussian matrix elements only slightly 
differs from the case of matrix elements taken from a flat 
distribution between $-1$ and $1$.

\begin{figure}[tbp] 
\includegraphics[width=8.0cm]{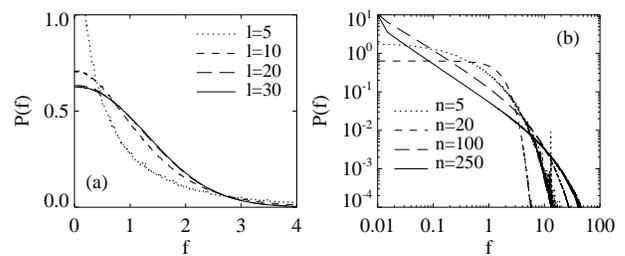} \centering 
\caption{(a) Force distributions obtained with $30\times 30$ random 
matrices (with pressure constraint), with increasing sparseness. The 
distributions for $l_z=30$ and $l_z=20$ are indistinguishable, but 
for smaller $l_z$ we see that the distribution becomes broader. (b) Here we 
show, for fixed $l_z=5$, the emergence of a power-law in $P(f)$ for 
large, sparse $n\times n$ matrices.} 
\label{fig.randmat3} 
\end{figure} 
 
{\em Sparse matrices} -- 
A property of real packing matrices that is not represented by the 
random matrices is their sparseness: only those forces that push 
directly onto a given particle contribute to the force balance, 
and hence, most matrix elements are zero.  
On average, each row contains $z$ nonzero elements, where $z$ 
denotes the average coordination number. 
In order to investigate whether this sparseness is responsible 
for the non-monotonic $P(f)$, we have generated a simple class of 
sparse random matrices: The matrices used are 
again $n\times n$, but now with only $l_z$ nonzero (Gaussian) 
elements per row (again, we leave the elements of the pressure contraint 
unaltered). These nonzero elements are arranged in a band-matrix like 
form. 
 
Force distributions for $n=30$ and several values of $l_z$ can be 
seen in Fig.~\ref{fig.randmat3}a. The maximum value of the 
distributions remains at $f=0$ and, surprisingly, it even 
increases as the matrix is more sparse. Uniformly distributed 
elements gave almost identical results. It thus appears that the 
characteristic peak of $P(f)$ is not directly related to the 
sparseness of the matrix. In addition we found that for large 
sparse matrices, the tail of $P(f)$ develops power-law scaling 
(Fig.~\ref{fig.randmat3}b).  
 
This demonstrates that a wide range of  
force distributions can be obtained by varying the matrix properties,  
and that there is no simple answer to the question what properties  
of the matrix ${\cal A}$ are necessary to mimic realistic packings. 
In the light of this discussion, let us make the 
following remark. Recently, Ngan \cite{ngan} obtained a variety 
of force distributions similar to those obtained for real packing 
matrices in Sec.~\ref{sect.numsnooker}, and compatible with the form of 
Eq.~(\ref{gaussmatrixresult}). 
These have been derived by minimizing an entropy functional under a 
pressure constraint similar to the one used in this paper \cite{footngan}, 
but without specifying the local microscopic equations of force balance. 
One may therefore wonder whether it is possible to make a connection 
between the force ensemble and Ngan's work. 
On the other hand, the results of this section clearly illustrate 
that properties of the local equations, which are absent 
in Ref.~\cite{ngan}, {\em do} play a crucial role: 
it can change $P(f)$ from Gaussian to power-law.

\subsection{Perturbing a physical packing matrix} 
 
\begin{figure}[tbp] 
\includegraphics[width=4.0cm]{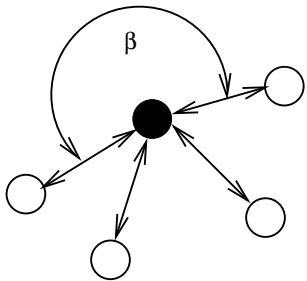} \centering 
\caption{Definition of a rattler. The net force on this rattler 
can only be zero if all forces involving this particle are zero. 
This means that the maximum angle between bonds, $\beta$, is 
larger than $\pi$. Such rattlers can arise when bonds are deleted  
or when the contact angles are randomly rotated (see text).} \label{fig.defrat} 
\end{figure}

In the previous section, we have shown that introducing elements from 
real packing matrices to random square matrices does not easily lead 
to the characteristic peak in the resulting $P(f)$.  Therefore, we now 
investigate the reverse route, i.e.  perturbing a real packing matrix 
by slowly introducing randomness in the matrix elements. We perform 
three sorts of perturbations. In the first, the angles of the contacts 
are randomly varied, which ensures that {\em the topology of the 
contact network remains unaltered}. In the second, we randomly delete 
contacts, in the third, we randomly add contacts. In all three cases, 
the $P(f)$ loses its maximum for sufficiently strong perturbation. 
We show how for the first two protocols, this behavior appears to be  
correlated to the emergence of ``rattlers'' (see Fig.~\ref{fig.defrat}). 
 
We have first constructed a matrix corresponding to an irregular 
packing of $1024$ bi-disperse disks (50:50 mixture, size ratio $1.4$) 
by molecular dynamics simulations using a 12-6 Lennard-Jones potential 
with the attractive tail cut off \cite{liuletter,PRL}.  This system is 
quenched below the glass transition ($k_B T_g \approx 1.1$) and its 
energy is minimized using a steepest descent algorithm, which 
guarantees that there is at least one stable force network. The 
resulting packing consists of $2814$ bonds so $z\approx 5.5$. 
 
\begin{figure}[tbp] 
\includegraphics[width=8.0cm]{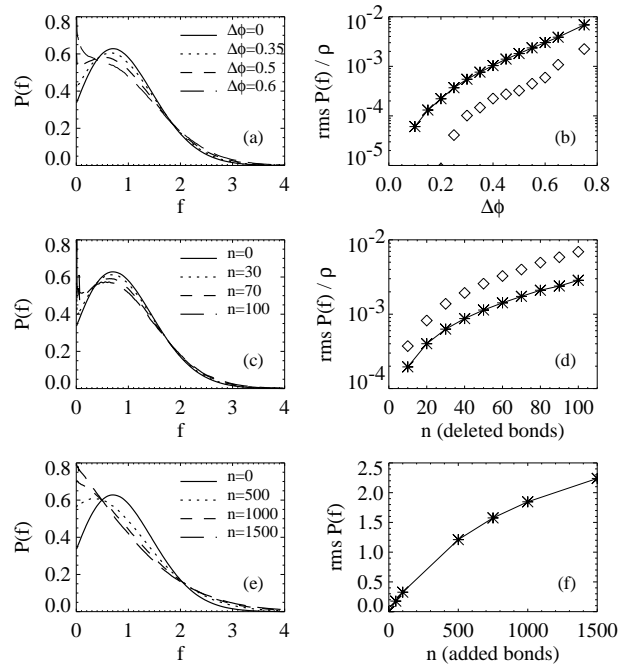} \centering 
\caption{Variation of $P(f)$ and number of rattlers when 
perturbing a realistic packing matrix. (a-b) Variations of 
the contact angle randomly selected from $[-\Delta \phi, \Delta \phi ]$;  
$P(f)$ evolves from peaked to monotonic (a). 
The density of rattlers $\rho$ (open symbols), and the rms variation of 
$P(f)$ (stars) with respect to the unperturbed situation 
are roughly proportional (b). A similar scenario occurs when bonds 
are randomly deleted (c-d). When bonds are added, however, no 
rattlers are created but $P(f)$ still evolves to a monotonic form 
(e-f).} \label{newperturb} 
\end{figure}

The effects on the $P(f)$ for the force ensembles corresponding to the 
perturbed matrices is illustrated in Fig.~\ref{newperturb}. In 
Fig.~\ref{newperturb}a-b we illustrate the effect of rotating the 
contacts by random angles uniformly generated between $-\Delta\phi$ 
and $\Delta\phi$. With increasing $\Delta\phi$, the packing is 
getting more and more unphysical (corresponding less and less to a 
system of non-overlapping particles). Nevertheless, the topology of 
the network always remains the same, and Newton's third law is always 
respected. The resulting force distributions are computed using the algorithm 
described in Sec.~\ref{sec.snooker} and averaged over all randomly 
generated perturbations of our original matrix $\cal{A}$.  
In Fig.~\ref{newperturb}a we have plotted $P(f)$ for different  
values of $\Delta\phi$. For small $\Delta\phi$ we obtain the 
characteristic shape of $P(f)$ for jammed systems similar to 
Fig.~\ref{fig.fig2}. Small perturbations ($\Delta\phi < 0.2$ rad) 
hardly change $P(f)$, but at larger $\Delta\phi$ the peak around 
$\left\langle f \right\rangle$ disappears and $P(f)$ looks 
``unjammed''.  For $\Delta\phi > 0.75$ we were no longer able to 
obtain solutions with all $f_j \geq 0$.

This clearly shows that the conditions of a sparse 
matrix respecting the packing topology, elements distributed 
between $[-1,1]$, and the incorporation of Newton's third law into 
$\cal{A}$ are not sufficient to obtain the characteristic peak in 
$P(f)$. Even at relatively small perturbations of $\cal{A}$ the 
shape of $P(f)$ changes quite abruptly. Furthermore, our 
simulations clearly show that we are not even guaranteed to find a 
solution of the problem for a randomized matrix: only a very small 
fraction of all possible matrices lead to a solution for which all 
$f_j \geq0$. So, even though the emergence of a non-monotonic 
$P(f)$ is extremely robust for packing matrices, it appears to be 
not at all a generic feature for arbitrary matrices. 
 
The amount of rattlers (Fig.~\ref{fig.defrat}) due to the randomization  
of the angles is small,  
but can be seen as a crude measure of the contact geometry. To 
our suprise, the evolution of the average amount of rattlers, and the 
rms deviation of $P(f)$ from the unperturbed distribution are fairly 
proportional (Fig.~\ref{newperturb}b). Here, this rms deviation has been  
measured as $\sqrt{\int df (P_0(f)-P(f))^2}$,  
where $P_0(f)$ denotes the unperturbed distribution. 
 
When bonds are deleted, a similar scenario occurs.  Again the 
$P(f)$'s loose their peak and the rms deviation of $P(f)$  
follows the amount of rattlers quite well (Fig.~\ref{newperturb}c-d).  
On the other hand, when bonds are 
added, no rattlers are generated, but the $P(f)$ still exhibit the 
same trend (Fig.~\ref{newperturb}e-f). Curiously, all the $P(f)$'s 
for the cases of added contacts appear to intersect in two points 
(Fig.~\ref{newperturb}e); we have no explanation for this 
phenomenon.

\section{Discussion}\label{sec.discussion} 
 
In this paper we have proposed a novel ensemble approach to 
athermal hard particle systems. The full set of mechanical 
equilibrium constraints were incorporated, in contrast to more 
local approximations or force chain models  
\cite{qmodel,edwards3D,ngan,max_entropy,halseyinagroove,chainmodel}.  
The basic idea is to exploit the separation of force and 
packing scales by simply averaging with equal probability over all 
mechanically stable force configurations for a fixed contact geometry.  
There are thus two important ingredients, namely the assumption of a 
flat (Edwards-like) measure in the force space and the fact that 
packings are hyperstatic. As the flat or uniform measure can not 
be justified from first principles, the emerging force probability 
distribution $P(f)$ provides a first important test. For small 
forces, the ensemble nicely reproduces the typical non-monotonic 
behavior that has been found in numerous experiments and numerical 
simulations. Also, $P(f)$ remains finite at $f=0$, which has been 
the problem of earlier models \cite{edwards3D,qmodel}. 
 
\begin{figure}[tbp] 
\includegraphics[width=8.0cm]{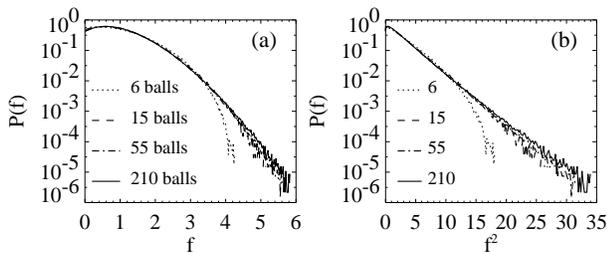} \centering 
\caption{Logarithmic plots of the $P(f)$ for snooker triangles of 
increasing size as function of $f$ (a) and $f^2$ (b) illustrate 
that the tails of these distribution decay faster than exponential 
but slower than Gaussian.} \label{fig.tails} 
\end{figure} 
 
Let us now discuss the tails of the distribution.  From 
Eq.~(\ref{pfresult}) we can only predict the asymptotic behavior of the 
slowest decaying term, corresponding to the minimal value of 
$b_\lambda$.  For 2D packings one can show that this minimal  
$b_\lambda\propto 1/\sqrt{N}$, but since $D\propto N$,  
the contribution to $P(f)$ of this term decays as $e^{-\sqrt{N}f}$;  
this term thus provides a sharp cutoff close to the maximal force.  
However, there will be a  
distribution of $b_\lambda$'s, and in order to resolve the tail of 
$P(f)$ one would really have to know all coefficients in 
Eq.~(\ref{pfresult}) for large enough systems.  In 
Fig.~\ref{fig.tails} we again plot the numerically obtained $P(f)$ 
for snooker-packings.  Although the systems are of limited size, 
it appears that the distributions have tails that neither are 
purely exponential nor purely Gaussian. The differences are 
subtle, and may be sensitive to numerical details. Even though 
numerical and analytical distributions for small packings appear 
to be in perfect agreement on a linear scale, on similar log 
scales the numerical curve seems to slightly underestimate the 
large fluctuations. While the numerical precision is about $10^{-6}$ around 
$\langle f \rangle$, the relative differences between numerical and exact 
results become about $5\% $ around $f = 4 \langle f \rangle$. 
In the literature, there has recently been 
some debate on the true nature of the tails \cite{taildiscussion}:  
while the carbon paper experiments undoubtedly yield exponential tails, it 
appears that most numerically obtained $P(f)$ display some downward 
curvature when plotted on log-lin scales.  It has also been argued 
that individual packings are not self-averaging and that tails 
appear Gaussian or exponential depending on how the ensemble is 
normalized \cite{epitomy}. At present, we can therefore 
neither confirm nor falsify the validity of the flat measure based 
on the tail of $P(f)$. 
 
Unger {\em et al.} \cite{Unger} recently proposed another test for the 
uniform measure. For frictional packings, they compared force  
configurations that emerge in a dynamical process to those obtained  
from a random sampling of the force space.  
They found that the dynamical solutions are located 
more centrally within the force space, and therefore concluded that 
the flat measure does not apply.  While this is definitely an 
interesting observation, this claim strongly depends on the ``flatness'' 
of their numerical sampling of the solution space, for which no 
evidence is provided. Counterintuitively, if the physical force 
networks were indeed more central, the ensemble $P(f)$ would even {\em 
overestimate} the large force fluctuations. Therefore, the validity 
of the flat measure remains an open issue. 
 
A second important ingredient of the force network ensemble is 
that there is no unique force solution for a given contact 
network, i.e. packings are hyperstatic.  While most packings are 
indeed hyperstatic, the precise degree of indeterminacy may depend 
on material parameters and construction history  
\cite{grestcoord,Unger,kasahara}.  It 
appears that strict isostaticity is only found for infinitely hard 
particles without friction, or with unphysically large friction 
coefficients. The present study was performed with highly 
coordinated regular packings, which are more hyperstatic than most 
physical packings.  The coordination number is therefore an 
important parameter that remains to be explored. It may very well 
be that the predictive power of the ensemble depends on this 
degree of indeterminacy. 
 
{\em Metabolic networks} -- While preparing this paper, we have 
become aware of a striking analogy between the force ensemble and 
the problem of metabolic networks \cite{otherbio,nature}.   
These are networks of biochemical reactions, in which the metabolite 
concentrations (particle positions) and the reaction and transport fluxes 
(interparticle forces) are the variables of the problem.  
In principle the fluxes follow from the concentrations, similar to 
how the forces follow from the particle positions. This coupling 
involves intricate reaction-diffusion dynamics, for which 
numerical values of most rates are not known.  In 
practice, however, a separation of time-scales occurs: the 
metabolite concentrations quickly adjust (seconds) to global 
changes in the network (minutes) \cite{timescales}. Very much 
like we employed the separation of length-scales, a successful 
strategy has been to treat the fluxes as independent variables and 
resolve the steady state from the stoichiometry of the network. 
 
Mathematically, the problem then reduces to an underdetermined 
matrix problem with non-negative flux variables, which is identical to 
the equations defining the force network ensemble.  
It turns out that for different metabolic maps the number of fluxes is  
always larger than the number of metabolites and therefore these systems  
are ``hyperstatic''.  
The main difference with respect to the force problem, however,  
is the network structure defining the matrix: 
metabolic networks are scale-free, i.e. with highly uneven 
connectivities.  This leads to power-law flux distributions 
\cite{nature}, which is very different from the $P(f)$ within the 
force ensemble.  This touches upon the deep question of how 
network statistics relate to the underlying network structures. 
In Sec.~\ref{sec.rm} we have found that, indeed, $P(f)$ can  
range from Gaussian to power-law when changing the properties of  
the matrix defining the ensemble. 
 
For metabolic networks the main interest is to find solutions in which 
the production of ``biomass'' is optimized.  In contrast to the 
averaging procedure within the force ensemble, this corresponds to 
finding the ``extreme pathways'' that form the corners of the 
hyperpolygon defining the solution space \cite{otherbio}.   
In fact, the force network solutions shown in  
Figs.~\ref{fig.3ballmode}  
and~\ref{fig.6ballforces}-\ref{fig.9ballsperiodic} are such 
extreme pathways. 
We speculate that a systematic analysis of extreme solutions 
may give additional insight in the geometrical properties of the 
phase space and the emergent force statistics -- see also 
Ref.~\cite{herrmannensemble}. 
It would furthermore be interesting to see whether for disordered 
packings there still exist localized linearly propagating solutions 
such as shown in Fig.~\ref{fig.6ballforces}, 
or whether all particles have to be involved into the force network.

{\em Outlook} -- A number of crucial questions can possibly be 
addressed within our framework.  {\em (1)} By separating the 
contact geometry from the forces, we can start to disentangle the 
separate roles of contact and stress anisotropies in sheared 
systems. In particular, we have already shown that the ensemble 
comprises an ``unjamming'' transition for shear stresses above a 
critical value \cite{PRL}.  Furthermore, the contact and force 
networks exhibit different anisotropies under different 
construction histories \cite{vanel,2Dbehringer}.  
We suggest that the contact 
network anisotropies may be sufficient to obtain the pressure dip 
under sand piles.  {\em (2)} As also illustrated by 
\cite{Unger,herrmannensemble}, our approach is perfectly suited to 
include frictional forces. While these forces are difficult to 
express in a force law, they are easy to constrain by the Coulomb 
inequality. 
 
{\bf Acknowledgements.} We are grateful to Sorin T\u{a}nase-Nicola, 
Alexander Morozov, 
Kees Storm  and Wim van Saarloos for numerous illuminating discussions.  
J.H.S. and W.G.E. gratefully acknowledge support from the physics foundation  
FOM, and M.v.H. support from science foundation NWO through a VIDI grant. 
 
\appendix 
 
\section{The Pressure Constraint}\label{app1} 

In this appendix we first show that the sum of all forces $\sum 
f_{ij}$ is constant for regular packings under a fixed external 
pressure.  This provides a conservation law similar to the 
conservation of total energy in the microcanonical ensemble. We 
therefore revisit the problem of partitioning a conserved quantity in 
the second part of this appendix. 
 
One can compute the stress from the contact forces ${\bf f}_{ij}$ as 
 
\begin{equation}\label{stress} 
\sigma_{\alpha\beta} = \frac{1}{V} 
\sum_{\{ij\}} \left( {\bf f}_{ij} \right)_\alpha 
\left( {\bf r}_{ij}\right)_\beta~, 
\end{equation} 
where $V$ represents the volume of the system \cite{footcoarsegrain}. 
The vector ${\bf r}_{ij}={\bf r}_i - {\bf r}_j$ denotes the 
interparticle distance, which for monodisperse particles of diameter 
$\tilde{d}$ always has $|{\bf r}|=\tilde{d}$. 
For packings of frictionless disks, one can therefore write 
 
\begin{eqnarray} 
\sigma_{xx} &=& \frac{\tilde{d}}{V} 
\sum_{\{ij\}} f_{ij} \cos^2 \varphi_{ij}~, \\ 
\sigma_{yy} &=& \frac{\tilde{d}}{V} \sum_{\{ij\}} f_{ij} \sin^2 \varphi_{ij}~, 
\end{eqnarray} 
where $\varphi_{ij}$ indicates the angle of the contact with repect 
to the horizonal $x$-axis. Taking the trace of the stress tensor, 
we find $\sigma_{xx}+\sigma_{yy} =\frac{\tilde{d}}{V} \sum f_{ij}$. 
So indeed, a constant pressure condition is equivalent to a constraint 
for the sum of all contact forces, at least for monodisperse packings. 
To a good approximation this remains valid for polydisperse 
packings, since in practice, the forces are uncorrelated to $|{\bf r}|$ 
so that $\langle |{\bf r}|\rangle$ can be taken out of the sum in 
Eq.~(\ref{stress}) \cite{radjai,PRL}. 
 
Let us now consider the statistical properties of a set of 
$n$ independent non-negative variables $x_j \geq 0$, that is contrained 
by a conservation law 
 
\begin{equation}\label{conservation} 
\sum_{j=1}^n x_j = X~. 
\end{equation} 
The phase space of these variables is a $(n-1)$ dimensional 
simplex of volume 
 
\begin{eqnarray} 
\Omega_n(X) &=& \left[ \prod_{j=1}^n \int_0^\infty dx_j \right] \, 
\delta \left(X - \sum_{j=1}^n x_j \right) \nonumber \\ 
&=& \frac{X^{n-1}}{(n-1)!}~, 
\end{eqnarray} 
where the integrals can be evaluated e.g. by Fourier representation 
of the $\delta$-function. 
Assigning an equal probability to all sets $\{ x_j\}$ obeying 
Eq.~(\ref{conservation}), we compute the probability of a single 
variable $P(x;n)$ as 
 
\begin{eqnarray}\label{px} 
P(x;n) &=& \frac{1}{\Omega_n(X)} 
\left[ \prod_{j=2}^n \int_0^\infty dx_j \right] \, 
\delta \left(X - x - \sum_{j=2}^n x_j \right) \nonumber \\ 
&=& \frac{\Omega_{n-1}(X-x)}{\Omega_n(X)} \,\Theta(X-x)\nonumber \\ 
&=& \frac{n-1}{X^{n-1}}\left( X-x\right)^{n-2} \,\Theta(X-x)~. 
\end{eqnarray} 
In the thermodynamic limit $n\rightarrow \infty$, this becomes the 
purely exponential ``Boltzmann'' distribution 
 
\begin{equation} 
P(x;\infty) = \frac{1}{\langle x \rangle} \,e^{-x/\langle x \rangle}~, 
\end{equation} 
where $\langle x \rangle = X/n$. For finite $n>2$, however, this 
distribution is always monotonically decreasing. 
 
In this paper we encounter two (small) packing configurations for 
which the force network ensemble can be reduced to the simple 
problem discussed above, so that force distributions of the type 
Eq.~(\ref{px}) are found -- see Figs.~\ref{fig.3ballspace}  
and~\ref{fig.4ballsperiodic}.   
In general, however, the constraints of 
force balance on each particle are more complicated and lead to 
non-monotonic $P(f)$.

\section{Derivation of the Gaussian Random Matrix $P(f)$}\label{app2} 
In this appendix we show how Eq.~(\ref{gaussmatrixresult}) is obtained. 
We study the ensemble defined by 
 
\begin{eqnarray}\label{measure} 
P\left(\vec{f},{\cal A} \right) &=& 
\frac{1}{\Omega} \,P_a\left({\cal A}\right) 
\,\delta\left(F - \sum_{j=1}^n f_j \right)\nonumber\\ 
&&\times\prod_{i=2}^m \, \delta\left(\sum_{j=1}^n a_{ij} f_j \right)~, 
\end{eqnarray} 
where we define 
 
\begin{equation} 
P_a\left({\cal A}\right) = \prod_{i=2}^m \prod_{j=1}^n P_a(a_{ij})~. 
\end{equation} 
In order to be consistent with the notation in Sec.~\ref{sec.general}, 
we reserve the index $i=1$ for the inhomogeneous pressure constraint. 
The force distribution $P(f_j)$ becomes 
 
\begin{eqnarray}\label{pf} 
P(f_j) = \int d{\cal A} \, \prod_{k\neq j} \int_0^\infty df_k \, 
P\left(\vec{f},{\cal A} \right)~, 
\end{eqnarray} 
where

\begin{equation}
\int d{\cal A} = \prod_{i=2}^m \, \prod_{j=1}^n 
\int_{-\infty}^{\infty} da_{ij}~.
\end{equation}
The advantage of taking Gaussian elements $a_{ij}$ 
is that they can be integrated out explicitly, using the 
Fourier representations of Eq.~(\ref{fourierdelta}): 
 
\begin{eqnarray}\label{defpeff} 
&&\int d{\cal A} \, P_a\left({\cal A}\right) \prod_{i=2}^m 
\delta \left( \sum_{j=1}^n a_{ij}f_j \right) = \nonumber \\ 
&& \prod_{i=2}^m \int_{-\infty}^{\infty} \frac{ds_i}{2\pi} \, 
\prod_{j=1}^n \int_{-\infty}^{\infty} da_{ij} \,P(a_{ij}) \, 
e^{-i s_i\,a_{ij}f_j} = \nonumber \\ 
&& \prod_{i=2}^m \int_{-\infty}^\infty \frac{ds_i}{2\pi} \, 
e^{-\frac{1}{4} s^2 \sum_j f_j^2} = 
\left( \frac{1}{\pi \sum_j f_j^2}\right)^{(m-1)/2}~. 
\nonumber \\ 
&& 
\end{eqnarray} 
It is convenient to bring the factor $\sum_j f_j^2$ to the exponent 
using the relation 
 
\begin{equation} 
\frac{1}{a^k} = \frac{1}{\Gamma(k)} \int_0^\infty dt \,t^{k-1} e^{-ta}~. 
\end{equation} 
Introducing this auxilary variable $t$, $P(f_j)$ becomes 
 
\begin{eqnarray}\label{horrible} 
P(f_j) &=& \frac{c}{\Omega} \prod_{k\neq j} \int_0^\infty df_k \, 
\delta\left(F-\sum_j f_j\right) \, \nonumber \\ 
&& \times \int_0^\infty dt t^{\frac{m-3}{2}} 
\, e^{-t \sum_j f_j^2} \nonumber \\ 
&=& \frac{c}{\Omega} 
\int_{-\infty}^\infty \frac{ds_1}{2\pi}\, e^{is_1(F-f_j)} 
\int_0^\infty dt \, t^{\frac{m-3}{2}} e^{-tf_j^2} \nonumber \\ 
&& 
\times \,\left[g(is_1,t)\right]^{n-1}~, 
\end{eqnarray} 
where we have used a shorthand 
 
\begin{equation} 
g(is_1,t)=\int_0^\infty df \, e^{-is_1 f-t f^2}~. 
\end{equation} 
We now exponentiate the full integrand of Eq.~(\ref{horrible}), 
so that $P(f_j)$ becomes 
\begin{equation}\label{nare_integraal} 
P(f_j) = \frac{c}{\Omega} 
\int_{-\infty}^\infty \frac{ds_1}{2\pi} \int_0^\infty dt \, 
e^{-is_1f_j -tf_j^2}\,e^{\Phi(is_1,t)}~, 
\end{equation} 
where 
\begin{eqnarray} 
\Phi(is_1,t) &=& is_1F + \left(\frac{m-3}{2}\right)\ln(t) \nonumber \\ 
&& + (n-1) \ln\left[ g\left(is_1,t\right)\right]~. 
\end{eqnarray} 
If we now fix $\langle f\rangle =1$ by taking $F=n$, one observes 
that all terms of the phase $\Phi$ are extensive in $n$ or $m$. 
In the limit where both $n,m\rightarrow \infty$, we can thus 
evaluate the integrals using a saddle-point approximation. 
By determining the stationary phase, i.e. 
$\partial\Phi/\partial s_1 =0$ and $\partial \Phi/\partial t=0$, 
one finally arrives at the result of Eq.~(\ref{gaussmatrixresult}): 
\begin{equation}\label{gaussmatrixresultappendix} 
P(f) = c(\rho) \, e^{-(1-\rho) f} \, e^{-b(\rho)f^2}~. 
\end{equation} 
The function $b(\rho)$ varies almost linearly between $b(0)=0$ and  
$b(1)=1/\pi$.


\begin{thebibliography}{99}
\bibitem{footfacets} For the problem of partitioning 
$E_{\rm tot}$ into $n$ positive energies $E_i$, the phase space is 
a $n-1$ dimensional simplex that is bounded by 
$n$ facets -- see appendix~\ref{app1}. In the thermodynamic limit 
the ratio $n_{\rm facets}/D \rightarrow 1$, leading to the 
Boltzmann distribution $e^{-\beta E_i}$. In the force network 
ensemble this ratio $n_{\rm facets}/D = z/(z-2d)>1$, which 
explains the suppresion of $P(f)$ for small forces with respect to 
a pure exponential.
\bibitem{footlorentz} We have also been able to solve the problem 
for a Lorentzian $P_a(a)\propto 1/(1+a^2)$. 
Independent of the value of $\rho$, we surprisingly find that this extreme 
distribution always leads to $P(f)=e^{-f}$.
\bibitem{footngan} Note that in Ref.~\cite{ngan} a somewhat different 
pressure constraint is employed, since variations in the coordination 
number may lead to variations in $\langle f \rangle$ at a fixed pressure.
\bibitem{footcoarsegrain} We sidestep here the subtle issue of the coarse 
graining necessary to derive a macroscopic stress field from a microscopic 
particle model. For a detailed discussion on this, see e.g. \cite{goldhirsch}.
\end{thebibliography}
\end{document}